\documentclass[onecolumn,preprint,showpacs,preprintnumbers,amsmath,amssymb,superscriptaddress]{revtex4-1}
\usepackage{graphicx}
\usepackage{dcolumn}
\usepackage{bm}
\usepackage{color}

\usepackage[english]{babel}
\usepackage[utf8x]{inputenc}
\usepackage[T1]{fontenc}

\usepackage{amsmath}
\usepackage[colorinlistoftodos]{todonotes}
\usepackage[colorlinks=true, allcolors=blue]{hyperref}

\begin{document}

\title{Multiphase imaging of freezing particle suspensions by confocal microscopy}
\author{Dmytro Dedovets}
\affiliation{Ceramic Synthesis \& Functionalization Lab, UMR3080 CNRS-Saint-Gobain, 84306 Cavaillon, France}
\author{Sylvain Deville}
\email{sylvain.deville@saint-gobain.com}
\affiliation{Ceramic Synthesis \& Functionalization Lab, UMR3080 CNRS-Saint-Gobain, 84306 Cavaillon, France}

\date{\today}

\begin{abstract}

Ice-templating is a well-established processing route for porous ceramics. Because of the structure/properties relationships, it is essential to better understand and control the solidification microstructures. Ice-templating is based on the segregation and concentration of particles by growing ice crystals. What we understand so far of the process is based on either observations  by optical or X-ray imaging techniques, or on the characterization of ice-templated materials. However, in situ observations at particle-scale are still missing. Here we show that confocal microscopy can provide multiphase imaging of ice growth and the segregation and organization of particles. We illustrate the benefits of our approach with the observation of particles and pore ice in the frozen structure, the dynamic evolution of the freeze front morphology, and the impact of PVA addition on the solidification microstructures. These results prove in particular the importance of controlling both the temperature gradient and the growth rate during ice-templating.

\end{abstract}
\maketitle

\section{Introduction}

Ice-templating is a well-established processing route for porous materials in materials science in general~\cite{Zou2010,Ahmed2011,Qiu2012,Walter2013,Pawelec2014b,Naviroj2015} and in ceramics in particular~\cite{Tang2015a,Ferraro2017,Rachadel2017}. Hundreds of papers are now published every year on the topic, and the structural or functional properties of ice-templated materials are systematically explored~\cite{Deville2017h}. Although a considerable number of applications have been proposed, moving ice-templating from the lab towards industrial applications will depend, to a large extent, of our ability to finely understand and control the process to ensure reproducible, reliable architectures. 

Ice-templating is based on the segregation of matter by growing crystals, which can then be concentrated between the later. Removal of the ice--whereby ``ice'' is a generic term for crystals grown from the solvent--provides a macroporous scaffold where the pores are a replica of the ice crystals, and the organization of particles in the scaffold is obtained during freezing. The structure of ice-templated materials, and thus their properties, are therefore largely controlled by the phenomena that takes place during freezing. 

A lot of attention has thus been paid to in situ observations of the freezing of particle suspensions. Several techniques have been proposed, each having its advantages and limitations~\cite{Deville2017c}. Most of what we understand from the interactions of particles with growing crystals have been obtained by optical microscopy~\cite{Korber1988,Saint-Michel2017}. However, it only provides 2D observations, and the spatial resolution is not sufficient when small particles are used. X-ray imaging can provide 3D reconstruction of the grown~\cite{Deville2009b,Delattre2014} or growing~\cite{Deville2013} crystals. However, its spatial resolution is not sufficient either to image particles. Artifacts induced by the beam are also still problematic~\cite{Deville2013}. Transmission electron microscopy has also been used~\cite{Tai2014} but does not provide 3D observations, and is not appropriate for systematic experiments. As the sample is fixed and of small dimensions, only a few interactions events can be imaged.

In absence of appropriate experimental observations, efforts have been put in modelling the redistribution of particles by growing crystals. Discrete elements modelling, in particular, provided numerous insights into the physics of ice-templating. The role of the growth rate of the crystals on the ordering of monodispersed spherical particles~\cite{Barr2010} or the alignment of anisotropic (platelets) particles~\cite{Bouville2014d}, was assessed. These results can be used experimentally to produce a variety of materials with controlled microstructures and functional or structural properties~\cite{Kim2009}. Further progress in our understanding now depends on in situ observations of these phenomena.

Ideally, we need thus a technique able to image in situ and without artifacts the growth of ice crystals and the redistribution of particles, as well as the later stages of freezing when ice invades the pores between concentrated particles. We recently demonstrated how confocal microscopy could be used to image in situ in 3D the growth of ice crystals~\cite{Marcellini2016}. Here we built on this preliminary work and show that particles can also be imaged during freezing. We developed a cooling stage that provides an independent control of the temperature gradient and the growth rate of the ice crystals, making systematic investigations possible. In this paper, we demonstrate the benefits of this approach to investigate the physics of ice-templating.

\section{Methods}

We used 2~$\mu m$ diameter PMMA/TEFMA particles, marked in fluorescence with Pyrromethene 546 (emission wavelength: 519~nm). The fluorescent dye (Sulforhodamine B) is dissolved in water at $10^{-4}M$. The suspensions were then prepared by incorporating 1~vol.\% of particles in this aqueous solution. The particle concentration is lower than that typically used in ice-templating (5--40~vol.\%) because we cannot properly image a volume if the particle concentration is too high, because of light scattering. Using index-matched particles would be more ideal for imaging, but this means not using pure water as index matching is usually obtained in water/DMSO systems. We prefer to use suspensions which formulation (aqueous) is close to that of ice-templating. The suspensions were sonicated in an ultrasound bath for a few minutes to ensure a good dispersion of the particles. A few suspensions were prepared by dissolving 1~wt.\% of PVA (POLYVIOL SOLUTION LL 2830, 25\%) in the suspension.

\begin{figure}
\centering
\includegraphics[width=8cm]{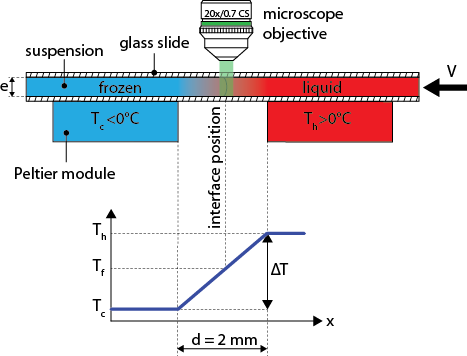}
\caption{Experimental setup to perform in situ freezing experiments in the confocal microscope. A constant temperature gradient $\Delta T = T_h - T_c$ is established in the gap $d$ between the Peltier modules. The sample is translated through the temperature gradient at a constant velocity $V$, which induces a growth of the ice crystals at a velocity $V$. The interface is thus kept at a constant position in the observation frame. Samples are 100~$\mu m$ thick. $T_f$ is the freezing point of the suspension. \copyright (2017) Sylvain Deville (10.6084/m9.figshare.5722675) CC BY 4.0 license https://creativecommons.org/licenses/by/4.0/}
\label{fig:figure0}
\end{figure}

We developed a cooling stage to perform in situ freezing experiments under the confocal microscope. The setup (Fig.~\ref{fig:figure0}) is composed of two Peltier modules that provide a constant temperature gradient in the gap $d$ between them. The particle suspension is introduced in a Hele-Shaw cell made of two glass slides, separated by two stripes of double-side sticky tape which act as spacers to ensure a constant sample thickness of 100~$\mu m$. The sample is sealed on both sides. This assembly is translated along the $x$-axis through the temperature gradient at a constant velocity $V$ by a stepper motor (Micos Pollux Drive stepper motor with VT-80 translation stage (PI, USA)). Because the sample is thin (100~$\mu m$), thermal equilibrium is achieved in the range of growth rate investigated here (1--50~$\mu m/s$); the solidification front is thus at a constant position in the observation frame. We can therefore vary independently the solidification front velocity (adjusted by the stepper motor) and the temperature gradient (established by the Peltier modules).

Confocal imaging is achieved by two different fluorophores: the dye incorporated in the particles, and a second dye (Sulforhodamine B) dissolved in water at $10^{-4}M$, which fluoresces at 586~nm. Experiments performed with and without Sulforhodamine B revealed that it has no noticeable impact on the freezing behavior of the system~\cite{Dedovets2017}. Images are acquired through two photodetectors, operating at the respective emission wavelengths of the dyes. For image acquisition, we used long working distance non-immersive objectives (Leica HC PL APO 20x/0.70 CS and 10x/0.40 CS2) to minimize the effect of the microscope thermal mass on the freezing process. These objectives have free working distances of 0.59~mm and 2.2~mm respectively. 

In a typical experiment, the sample is put in place on top of the Peltier elements, thermal insulation is achieved by covering the sample with a piece of polyurethane foam. A hole in the foam let the objective come in close contact to the sample. The desired temperature gradient $\Delta T$ is established by setting the temperatures of Peltier elements and the sample is then put in motion by the stepper motor. The interface velocity stabilizes within a minute after the beginning of the sample translation. The temperature gradient was varied from $5^\circ C/mm$ to $15^\circ C/mm$ in the experiments. The solidification front velocities were varied from 1~$\mu m/s$ to 40~$\mu m/s$.

Depending on the experiments and features investigated, 2D or 3D images were acquired. Image reconstruction was done with Fiji (ImageJ 1.51h)~\cite{Schindelin2012}.

\section{Results and discussion}

The benefits of confocal microscopy to investigate the ice growth and the segregation of particles in a suspension are four-fold: 
\begin{itemize}
	\item we can image individual particles. It is thus possible track the dynamics and organization of particles during and after freezing.
	\item because the dye is expelled from the growing ice, we can easily discriminate between the water and the ice phases on the images. We can thus image simultaneously the particles, the water, and the ice. 
	\item because of the rapid imaging mode of the microscope, we can image the process in 2D at a rapid time resolution: up to 40~Hz at $512 \times 512$ pixels. We can thus take snapshots of the solidification microstructures even at fast growth velocities (up to 40~$\mu m/s$ here.)
	\item we can use the confocal mode to reconstruct the 3D solidification microstructures.
\end{itemize} 

Such combinations provide unprecedented insights into the phenomenon investigated here, as we demonstrate below. 

\subsection*{Particle-scale observations}

\begin{figure}
\centering
\includegraphics[width=8cm]{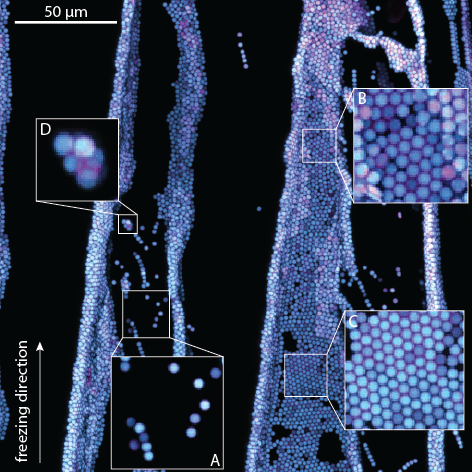}
\caption{3D reconstruction of the solidification microstructure. Although most of the structure is frozen, some residual liquid pockets can be observed between particles. Several features of interest are pointed out here: A: isolated engulfed particles. B: Pore ice in larger pores. The ice already invaded the larger inter-particle pores. C: Locally dense packing of particles: pore ice is moving through the packing, into the smaller pores. D: residual liquid pockets in a small agglomerate of particles. Growth rate: 20~$\mu m/s$, temperature gradient: $5^\circ C/mm$. \copyright (2017) Sylvain Deville (10.6084/m9.figshare.5722675) CC BY 4.0 license https://creativecommons.org/licenses/by/4.0/}
\label{fig:figure1}
\end{figure}

The observation of a partially-frozen structure (Fig.~\ref{fig:figure1}) already provides a fresh look at the organization of particles and the late stages of freezing. Elongated regions of concentrated particles, typically encountered in ice-templated materials, can be observed between the crystals. The directionality of growth induced a directional segregation pattern. 

If most of the particle reorganization took place here between adjacent ice crystals, some of the particles were also segregated between the crystal and the lower glass slide, resulting in the particle monolayer regions seen for instance in the insets B and C of Fig.~\ref{fig:figure1}. Although this is just a boundary limit effect, it provides an ideal configuration to investigate the penetration of ice in the dense packing of particles. Several configurations, highlighted by the different insets can be observed. In inset A, isolated particles were engulfed by the growing crystals. We could not find evidence of a liquid film around these particles. Although a premelted film is probably still present~\cite{Dash1999,Wettlaufer2006}, its expected thickness (a few nm) makes it too small to be observable by confocal microscopy. Inset B shows a partially frozen region. Local defects in particle packing can be observed, resulting in pores of different sizes between the particles. The pores between closely packed particles still contain liquid water (as seen by the fluorescence of the dye), while the water in the larger pores is frozen (no fluorescence visible anymore). This observation can be explained by the Gibbs-Thomson depression of the freezing point. The ice entry temperature is lower as the pore size diminishes~\cite{Wettlaufer1999a,Liu2003}. In inset C, which corresponds to a region at a lower temperature, pore ice can be observed in a densely-packed region. The upper part of the inset is mostly liquid, while the lower part is mostly frozen. The local temperature is therefore probably close to the ice entry temperature for the pore size that corresponds to the close packing of particles. The same behavior can be observed in isolated 3D agglomerates of particles (inset D). A group of particles were concentrated and engulfed by the ice, forming a small agglomerate. The fluorescence signal clearly indicates that liquid water is still present in the agglomerate. 

Fig.~\ref{fig:figure1} also shows that, for these solidification conditions, two fates are possible for the particles. Most of the particles are segregated by the growing crystals in the inter-crystal space. This mechanism is the basis for the development of ice-templated microstructures. However, some particles were also engulfed by the growing crystals without being concentrated. The particles are therefore isolated in the ice in the frozen structure. Such particles would not be observed if subsequent freeze-drying and sintering was performed, as they would just randomly fall into the structure upon removal of the ice. In situ observations of the frozen structures are therefore required to identify them. This means that the segregation of particles by the growing crystals is not 100\% efficient. Such behavior was not reported before, and knowing about it is important as this could affect the ice-templated microstructures. 

\begin{figure}
\centering
\includegraphics[width=8cm]{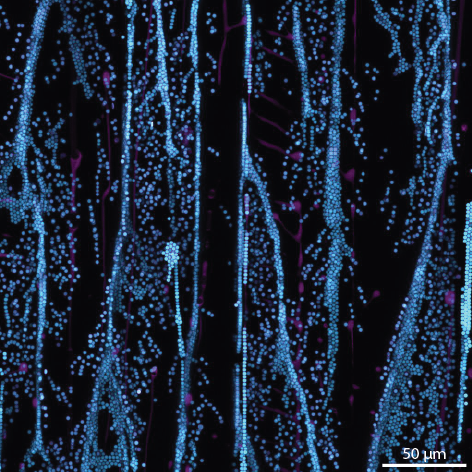}
\caption{3D reconstruction of the solidification microstructure. The engulfment of isolated particles is enhanced by the increase of growth rate from 20~$\mu m/s$ to 40~$\mu m/s$, compared to fig.~\ref{fig:figure1}. Growth rate: 40~$\mu m/s$, temperature gradient: $5^\circ C/mm$. \copyright (2017) Sylvain Deville (10.6084/m9.figshare.5722675) CC BY 4.0 license https://creativecommons.org/licenses/by/4.0/}
\label{fig:figure2}
\end{figure}

An increase of the growth rate to 40~$\mu m/s$ (for the same temperature gradient) enhances this behavior (Fig.~\ref{fig:figure2}). We can observe that the fraction of isolated, engulfed particles is increased. This can be explained by two phenomena: 
\begin{itemize}
	\item the increase of the growth rate favors the engulfment of particles. The rejection or engulfment of the particles by the freeze front is governed by the force balance on the particles~\cite{Rempel2001a}. In its most simple description, the force balance takes into account the viscous drag on the particle (exerted by the flux of water towards the interface) and the intermolecular interactions between the solidification front and the particle. When the drag is greater than the intermolecular interactions, the particle is engulfed by the front. There is therefore a critical velocity for engulfment at a given particle size or, conversely, a critical particle size at a given freeze front velocity. However, the particle-particle interactions modify the force balance, and favors the engulfment. In a concentrated system like here, the critical velocity for engulfment is therefore lower than for isolated particles. The most important aspect here is that the concentration of particles at the tip of the crystals will favor their engulfment.
	\item the increase of the growth rate destabilizes the interface, as illustrated later in the paper (see Fig.~\ref{fig:figure6}). Local variations of the ice growth velocity favors the engulfment of isolated particles.
\end{itemize}

Further work is now required to investigate the dependency of this phenomenon upon the freezing conditions (particle concentration, growth rate, temperature gradient). 

\subsection*{Particle organization induced by ice growth}

\begin{figure}
\centering
\includegraphics[width=8cm]{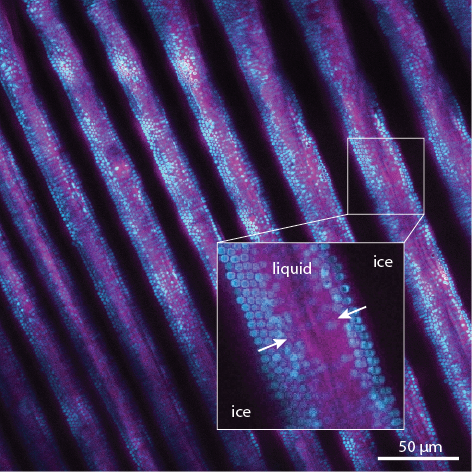}
\caption{2D cross-section that shows the organization of the particles induced by lateral growth of the crystals. The spatial organization can be compared with the output of discrete elements modelling. The first few layers of particles close to the crystal surface are organized. The particles further show less organization. 1~wt.\% of PVA, growth rate: 2~$\mu m/s$, temperature gradient: $10^\circ C/mm$. \copyright (2017) Sylvain Deville (10.6084/m9.figshare.5722675) CC BY 4.0 license https://creativecommons.org/licenses/by/4.0/}
\label{fig:figure3}
\end{figure}

The understanding and control of the organization of particles or more generally objects induced by the growth of crystal is a central question in solidification studies~\cite{Deville2017}. In particular, it is necessary to understand how the ordering, packing, and orientation of particles is related to the freezing conditions and morphology of the growing crystals. However, because of the typical space and time scale associated to the ice-templating conditions, these information’s were not accessible experimentally so far. Here, confocal microscopy is a suitable tool to investigate these phenomena. An horizontal cross-section taken during freezing is shown in Fig.~\ref{fig:figure3}. Parallel ice crystals (in black) repel and concentrate the particles in the inter-crystals space. The regions still liquid appear in magenta. The particles and their organization can clearly be observed in the regions adjacent to the crystals surface. In particular, we can see how the first 2--3 layers of particles, in direct contact with the growing crystals, organize into dense hexagonal packings. When the thickness of the layer of collected particles increases, the ordering is progressively lost.

These observations are in good agreement with the output of discrete elements modelling~\cite{Barr2010,Bouville2014d}, which predicted such segregation patterns. The particle organization was predicted to be strongly dependent on the displacement velocity of the interface in contact with the particles~\cite{Barr2010}. If a growth velocity of 2~$\mu m/s$ was set in this particular case, the lateral growth of the crystals, which is responsible for collecting and concentrating the particles, is much lower than this (less than 2~$\mu m/s$ in this case). Previous X-ray tomography performed in situ during freezing also reported~\cite{Deville2013} that the lateral growth velocity of ice crystals in a colloidal suspension is 2 to 3 times lower than the growth velocity along the temperature gradient. The lateral growth velocity is therefore probably less than 1~$\mu m/s$ here.

Confocal microscopy appears thus as a powerful tool to investigate particle reorganization during freezing. Following this initial assessment, future work will focus on systematic variations of the freezing conditions to investigate its impact on the reorganization of particles.

\subsection*{Effect of crystal tilt on particle segregation}

\begin{figure}
\centering
\includegraphics[width=8cm]{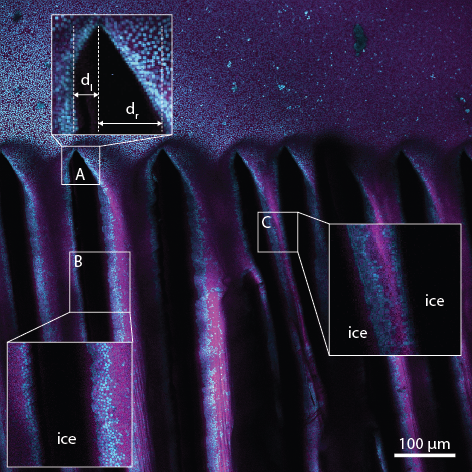}
\caption{Snapshot of the system during ice growth with titled crystals. A differential lateral segregation of particles is induced by the tilt. The tilted tip of the crystals pushes particles to the right-hand side of the crystals. A: the tip of the crystal is asymmetric, and its relative surfaces (or length, in 2D) that collect the particles on both sides are therefore different. B: Different concentration of particles are found on the right-hand side and left-hand side of the crystals, because of the tilt-induced uneven segregation. C: progressive concentration of particles induced by the lateral growth of the ice crystals. 1~wt.\% PVA, growth rate: 1~$\mu m/s$, temperature gradient: $10^\circ C/mm$. \copyright (2017) Sylvain Deville (10.6084/m9.figshare.5722675) CC BY 4.0 license https://creativecommons.org/licenses/by/4.0/}
\label{fig:figure4}
\end{figure}

Ice crystals, similar to other solidification systems such as succinotrile~\cite{Ghmadh2014}, do not always grow perfectly aligned with the temperature gradient. Depending on the growth rate, temperature gradient, and the presence of additives, ice crystals can grow titled. The tip of the crystals can, in such case, adopt an asymmetric morphology which may, in turn, impact the particle redistribution. An example is shown in Fig.~\ref{fig:figure4}, with a snapshot taken during the growth of ice. The asymmetric morphology can clearly be observed. The tip of the crystals is partly faceted, with a facet facing the incoming particles. The projected area (or length, in 2D) that collects the particles is shown in inset A. The projected length $d_r$ of the right-hand side of the crystal is greater than the projected length $d_l$ of the left-hand side of the crystal. Because of this tilt and so the different surfaces available to collect the particles, the relative fraction of particles segregated to the right-hand side of the crystals is greater than that segregated to the left-hand side. This can be observed with more details in the inset B, that shows both sides of the same crystal. The thickness of the layer of particles rejected by the lateral growth of the crystal is 2 to 3 times larger on the right-hand side than on the left-hand side. In a given pore channel, such as shown in inset C, the segregation of particles is thus asymmetric. 

This behavior was not reported before, and may impact the particle segregation pattern. The particle organization (packing density, orientation in case of anisotropic particles) between the crystals depends, among other parameters, on the thickness of the accumulated layer of particles. Several recent studies have reported that a controlled segregation of particles by ice crystals can be used to induce 2D or 3D structures with structural or functional properties~\cite{Kim2009,Shen2011,Bouville2014,Ferraro2016}. In such cases, the functional response may be optimized by a better control of the particle organization. The effect identified here could thus potentially be used to tune these properties. 

\subsection*{Impact of temperature gradient}

The temperature gradient, along with the growth velocity, is the parameter with the most dramatic impact on the solidification microstructure. However, in the majority of the ice-templating setups, the temperature gradient is not controlled and evolves as freezing proceeds. The usual approach in ice-templating is to put the suspension in a mold, the base of which is cooled at a constant cooling rate~\cite{Deville2017g}. In such a configuration, the temperature gradient varies during the experiment. Most studies have thus focused on the impact of the growth rate~\cite{Waschkies2009,Deville2006}, which is a much easier parameter to control.

Our temperature-controlled setup provides an independent control of the cooling rate and the temperature gradient. We can thus assess separately the impact of the later.

\begin{figure}
\centering
\includegraphics[width=18cm]{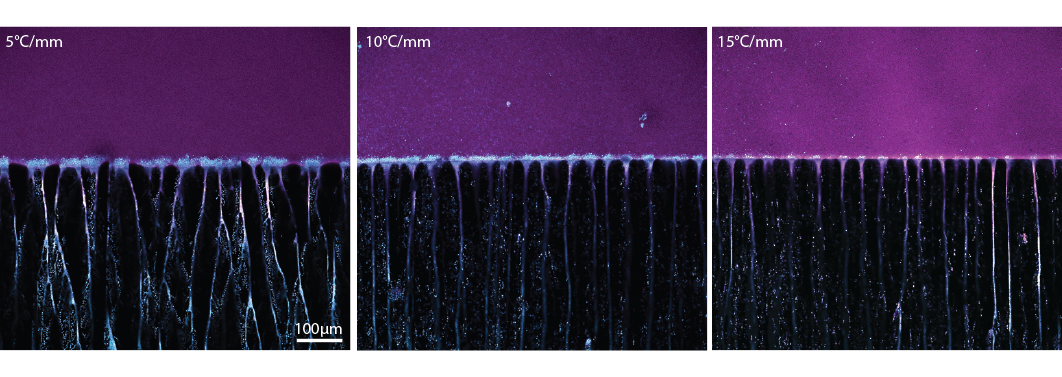}
\caption{Effect of the temperature gradient on solidification pattern and particle segregation at the tip of the crystals. Increasing the magnitude of the temperature gradient for a given growth velocity stabilizes the front. Even at such rapid velocity (40~$\mu m/s$), a thin layer of particles repelled by the crystal tip is visible. The corresponding temperature gradient is indicated on each picture. \copyright (2017) Sylvain Deville (10.6084/m9.figshare.5722675) CC BY 4.0 license https://creativecommons.org/licenses/by/4.0/}
\label{fig:figure5}
\end{figure}

We investigated the impact of the temperature gradient on the solidification microstructure for a constant growth velocity (Fig.~\ref{fig:figure5}). The temperature gradient was varied from $5^\circ C/mm$ to $15^\circ C/mm$, for a growth rate of 40~$\mu m/s$. Two features should be noticed here. 


We can observe in Fig.~\ref{fig:figure5} that even at such rapid velocity (40~$\mu m/s$), a thin layer of accumulated particles is found at the tip of the crystals. This was not reported before, and could explain some of the previously published results. When the tip of the crystals reach the top of the sample, the layer of concentrated particles will be retained in the frozen sample, and thus in the final ice-templated material. This is usually not a problem as the layer is only a few microns thick, and the top and bottom of the sample are often removed (cut) before being used. However, ice-templated thin films have recently gained traction for a number of applications such as dye-sensitized solar cells or solid oxide fuel cells~\cite{Ren2007,Sofie2007,Chen2012a,Rachadel2017}. The top layer of an ice-templated film that is only a few microns thick cannot be removed, and may impact the final functional properties. The presence of accumulated particles at the top of the film can be seen for instance in Fig.~2c and 2d of ref.~\cite{Wang2015c}.

\begin{figure}
\centering
\includegraphics[width=18cm]{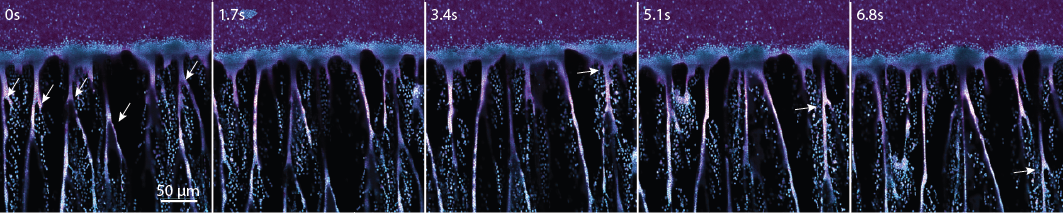}
\caption{Dynamic evolution of the crystal morphology, in an unstable regime. The solidification front in these conditions constantly evolve, with new crystals formed and older ones disappearing. This impacts the homogeneity of the solidification microstructure, but also results in a lot of isolated particles being engulfed. The arrows on the first frame indicate the tip of crystals that stopped growing. The arrows on the other frames indicate the same region in the successive frames; it illustrates how a grain boundary between adjacent ice crystals disappear from the freeze front and is replaced by another crystal tip. Growth rate: 40~$\mu m/s$, temperature gradient: $5^\circ C/mm$. \copyright (2017) Sylvain Deville (10.6084/m9.figshare.5722675) CC BY 4.0 license https://creativecommons.org/licenses/by/4.0/}
\label{fig:figure6}
\end{figure}

The second result is the stabilization of the crystals as the temperature gradient increases. When the crystals grow at 40~$\mu m/s$ in a $5^\circ C/mm$ temperature gradient, the crystals are not stable. A time-lapse sequence, shown in Fig.~\ref{fig:figure6}, reveals how the interface constantly evolves. New crystals appear at the front, while other ones disappear. This behavior has two consequences. The first one is the heterogeneous redistribution of particles. Ice-templating under such conditions would result in discontinuous pore channels, as a crystal that stops growing results in the closing of an ice-templated pore. The second consequence is the engulfment of a large fraction of isolated particles by the front, as shown in Fig.~\ref{fig:figure2} in the frozen structure. Again, from an ice-templating point of view, such behavior is not desirable. Increasing the temperature gradient from $5^\circ C/mm$ to $10^\circ C/mm$ and $15^\circ C/mm$, for the same growth velocity, stabilizes the front and should thus result in more homogeneous ice-templated architectures.

These results prove the importance of a proper control of both the growth rate and temperature gradient, because their impact on particle redistribution, but also on the homogeneity of crystals and thus of the ice templated structures. This should be kept in mind when designing ice-templating setups.

\subsection*{Effect of PVA}

Several additives are commonly used in suspensions for ice-templating. Additives include surfactants, but also binders, crystal growth modifiers, anti-foaming agents, or plasticizers. The impact of these additives on the development of ice-templated architectures is difficult to understand, as they typically affect several characteristics of the suspension (viscosity, surface tension, etc\ldots). With our setup, we can investigate the impact of additives on the morphology of ice crystals and the segregation of particles while keeping all other parameters constant. We illustrate here the approach with the impact of polyvinyl alcohol (PVA), which is one of the most used binder in ice-templating studies~\cite{Zuo2008,Pekor2010,Zuo2010,Yoon2010}.

\begin{figure}
\centering
\includegraphics[width=8cm]{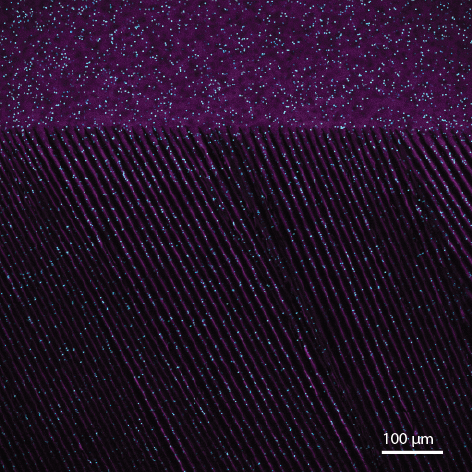}
\caption{Impact of PVA on the solidification microstructure. In presence of PVA, the crystals are tilted. The temperature gradient is vertical, the tilt is thus not an artifact that could be induced by a titled temperature gradient. Such a tilt is typically obtained in solidification system that exhibits a growth anisotropy. The crystals are also thinner, compared to the experiment without PVA. 1~wt.\% PVA, growth rate: 20~$\mu m/s$, temperature gradient: $10^\circ C/mm$. \copyright (2017) Sylvain Deville (10.6084/m9.figshare.5722675) CC BY 4.0 license https://creativecommons.org/licenses/by/4.0/}
\label{fig:figure7}

\end{figure}
The impact of the addition of 1~wt.\% of PVA in the particle suspension is shown in Fig.~\ref{fig:figure7}. At a growth rate of 20~$\mu m/s$, the crystals are titled with respect to the direction of the temperature gradient and solidification front. The tilt here is therefore a consequence of the PVA addition and is not an artifact due to a poorly controlled temperature gradient. A possible explanation is that the increased undercooling of the freeze front, resulting from the rejection of PVA by the growing ice, enhances the growth anisotropy of the ice crystals. This should be taken into account if PVA is added to prepare ice-templated materials, as tilted crystals will result in tilted pores in the final architecture. A tilt might not be desirable in applications that involve fluid or gas transport, for instance~\cite{Seuba2016c}.

\begin{figure}
\centering
\includegraphics[width=8cm]{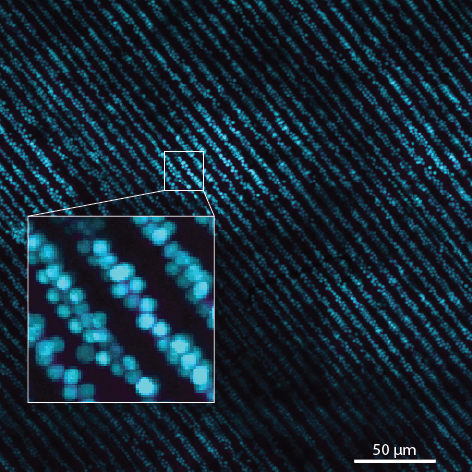}
\caption{2D grid of particles aligned in thin threads by the ice crystals. The initial suspension contained 1~wt.\% of PVA.  Growth rate: 20~$\mu m/s$, temperature gradient: $10^\circ C/mm$. \copyright (2017) Sylvain Deville (10.6084/m9.figshare.5722675) CC BY 4.0 license https://creativecommons.org/licenses/by/4.0/}
\label{fig:figure8}
\end{figure}

Crystals also become thinner with the addition of PVA. The periodicity of the lamellar solidification pattern in a $10^\circ C/mm$ is of 38~$\mu m$ (at 40~$\mu m/s$) and just 15~$\mu m$ (at 20~$\mu m/s$) with the addition of PVA (and should thus be even lower if we further increase the growth rate). The thinning of the ice crystals in presence of PVA can be used to align particles in 2D, as shown in Fig.~\ref{fig:figure8}. Here, the spacing between adjacent crystals is so thin that it can accommodate only 2 layers of particles. The particles used here sediment in a few minutes and thus rapidly rest at the bottom glass surface, before getting organized by the passing ice front. After freezing, we are thus able to obtain a 2D grid of particles aligned in thin threads. This has been used previously with $TiO_2$ and $VO_2$ suspensions~\cite{Romeo2014,Cao2014}, with applications in conducting platforms for electrical stimulation and thermochromic films. Our approach could thus be used to better understand and control such architectures.

\section{Conclusions and perspectives}

The 3D, multiphase imaging of confocal microscopy, combined with its rapid imaging capacities and submicronic spatial resolution, makes it the almost perfect tool to investigate the physics of ice-templating. Our specially designed temperature-controlled stage, with its independent control of growth rate of the crystals and temperature gradient, will now let us investigate in details the experimental parameters relevant to ice-templating, help us better understand the development of solidification microstructure, and guide us to adjust the experimental conditions based on the desired architecture and microstructure.  The results shown here also highlight the importance of a good control over the temperature gradient, to improve the stability of the freeze front and the homogeneity of the ice-templated architectures and microstructures.

\section*{Acknowledgements}
The research leading to these results has received funding from the European Research Council under the European Union's Seventh Framework Programme (FP7/2007-2013) / ERC grant agreement 278004 (project FreeCo). We acknowledge Tom Kodger for providing the particles used in this study.

\section*{References}

\bibliographystyle{model1-num-names}
\bibliography{biblio.bib}

\end{document}